\documentclass[twocolumn]{aastex62}
\graphicspath{{./}{figures/}}

\received{January 1, 2018}
\revised{January 7, 2018}
\accepted{\today}
\submitjournal{ApJ}

\shorttitle{Globular Clusters and Halo Masses}
\shortauthors{Burkert \& Forbes}

\begin{document}

\title{High-Precision Dark Halo Virial Masses from Globular Cluster Numbers: Implications for Globular Cluster Formation and Galaxy Assembly}

\correspondingauthor{Andreas Burkert}
\email{andi@usm.lmu.de}

\author{Andreas Burkert}
\affiliation{Universit\"ats-Sternwarte M\"unchen, Ludwig-Maximilians University Munich, University Observatory, Scheinerstr.\ 1, D-81679 M\"unchen, Germany}
\affiliation{Max-Planck Institute for Extraterrestrial Physics, Giessenbachstr.\ 1, D-85748 Garching, Germany}

\author{Duncan A. Forbes}
\affiliation{Centre for Astrophysics and Supercomputing, Swinburne University of Technology, Hawthorn VIC 3122, Australia}



\begin{abstract}
We confirm that the number of globular clusters (GCs), N$_{GC}$, is an excellent tracer of their host galaxy's halo virial 
mass M$_{vir}$. The simple linear relation M$_{vir} = 5 \times 10^9$ M$_{\odot} \times$ N$_{GC}$ fits the data perfectly 
from M$_{vir} = 10^{10}$ M$_{\odot}$ to M$_{vir} = 2 \times 10^{15}$ M$_{\odot}$. This result is independent 
of galaxy morphology and extends statistically into the dwarf galaxy regime with M$_{vir} = 10^8 - 10^{10}$ M$_{\odot}$, 
including the extreme ultra diffuse galaxy DF44. As this correlation does not depend on GC mass it is ideally suited for 
high-precision determinations of M$_{vir}$. The linearity is most simply explained by cosmological merging of 
a high-redshift halo seed population that hosted on average one GC per $5 \times 10^8$ M$_{\odot}$ of dark matter.  We show that hierarchical merging is also extremely powerful in restoring a linear correlation and erasing signatures of even a strong secular evolution of GC systems. The cosmological merging scenario also implies a strong decline of the scatter in $N_{GC}$ with increasing virial mass $\delta N_{GC}/N_{GC} \sim M_{vir}^{-1/2}$ in contrast with the observations that show a roughly constant scatter, independent of virial mass. This discrepancy can be explained if errors in determining virial masses from kinematical tracers and gravitational lensing are on the order of a factor of 2. GCs in dwarf satellite galaxies pose a serious problem for high-redshift GC formation scenarios; the dark halo masses of dwarf galaxies hosting GCs therefore might need to be an order of magnitude larger than currently estimated. 
\end{abstract}



\section{Introduction}
Dark halo virial masses are of essential importance in theories of galaxy evolution. They regulate galaxy dynamics, cosmic accretion rates, wind/outflow rates and tidal interactions. They also provide important information on the nature of dark matter. Historically, kinematical tracers such as rotation curves and the velocity dispersion of stellar and gaseous components have been used to measure the underlying dark matter distribution (for reviews see e.g. Sofue \& Rubin 2001,  Courteau et al. 2014 or Bland-Hawthorn \& Gerhard 2016). With baryons being usually more centrally concentrated than dark matter, these methods are mainly restricted to the inner regions. The outer halo parts can be investigated using satellite galaxies or gravitational lensing (Thomas et al. 2011; Merten et al. 2015). Still, this approach suffers from the limited number of satellite galaxies, projection effects, orbital anisotropy  and uncertainties in the 3-dimensional halo shape. 

It has been shown that globular cluster (GC) systems present another very promising approach to dark halo mass determinations. Blakeslee et al. (1997) found a linear correlation between the total number of GCs in a galaxy (N$_{GC}$) and its total mass for brightest cluster galaxies. Spitler \& Forbes (2009) identified an empirical near-linear relation between the stellar mass in a GC system (M$_{GC}$) and its host galaxy's dark matter virial mass (M$_{vir}$), over 4 orders of magnitude in halo mass. They used estimates of N$_{GC}$ from the literature and assumed a mean GC mass of $4 \times 10^5$ M$_{\odot}$. Total halo masses were estimated statistically from weak lensing studies. Strictly speaking, this work demonstrated a correlation between the number of GCs and M$_{vir}$. The correlation between GC system mass and virial mass has been subsequently extended by others, assuming different mean GC masses (Georgiev et al. 2010; Durrell et al. 2014) or a mean GC mass that varies with galaxy luminosity  (Hudson et al. 2014; Harris et al. 2017). 

For halo masses of log (M$_{vir}$/M$_{\odot}$)  $\ge$ 10.5, Harris et al. (2017) confirmed that M$_{GC}$ correlates with M$_{vir}$  with a power-law exponent fully consistent with unity. This was compared to using N$_{GC}$ for E/S0 galaxies only (e.g. excluding GCs associated with clusters of galaxies) for which they quote a non-linear slope of $\sim$0.9. However, the inclusion of GCs associated with clusters of galaxies, at log(M$_{vir}$/M$_{\odot}$) $\ge$ 14.5, would bring the slope closer to unity. They also examined the scatter of the two relations, finding a slightly tighter relation using directly observed N$_{GC}$ (i.e 0.26 dex) than using the derived M$_{GC}$ (i.e. 0.28 dex). Most recently, Forbes et al. (2018) extended an analysis to lower halo virial masses, using kinematically-based halo masses. They found that in this low mass regime GC systems were consistent with an extrapolation of the linear relation from higher masses (Spitler \& Forbes 2009). Thus the relation now extends over a remarkable 7 orders in halo mass. Forbes et al. also showed that only for the very lowest galaxy masses does N$_{GC}$ scatter about the relation more than using M$_{GC}$. 

Theoretical models seeking to reproduce the near-linear GC system-to-virial mass relation have included Kravtsov \& Gnedin (2005). Over the mass range 9.8 $<$ log(M$_{vir}$/M$_{\odot}$) $<$ 11.5 they predicted a M$_{GC}$ vs M$_{vir}$ slope of 1.13 $\pm$ 0.08. However, their simulations ended at z = 3.3 and so are not comparable to the relation observed at z=0. The recent semi-analytic model of El Badry et al. (2018) can reproduce the linear trend for halo masses log (M$_{vir}$/M$_{\odot})$ $\ge$ 12, but it progressively fails to reproduce the continuity of the M$_{GC}$ relation to the lowest virial masses. In their model, the predicted number of GCs formed is drawn from a power-law mass function with a minimum GC mass of 10$^5$ M$_{\odot}$.  Interestingly, they also introduce an alternative model of random mergers. This does a better job at reproducing the low mass regime than their semi-analytic model which includes various physical processes in a hierarchical cosmology. 

In this paper we further investigate the relation of M$_{vir}$ with N$_{GC}$ which is more closely related to the observations and much easier to measure than M$_{GC}$. In particular, we compare this relation to the expectations from hierarchical merging which provides a very simple and powerful explanation, independent of the complexity of GC formation. We demonstrate that within this scenario N$_{GC}$ is a high-precision indicator of dark halo virial mass which can be used in order to calculate current uncertainties in determining M$_{vir}$ from observations. The GC number -- virial mass relation also provides interesting insight into the origin of GCs, the role of high-redshift low-mass seed galaxies, the dark matter content of dwarf galaxies, constraints on GC disruption  and the importance of mergers in the growth of GC systems. 

\section{Observations}

Here we take data for N$_{GC}$ and M$_{vir}$ from the compilation of Spitler \& Forbes (2009) and supplement it with data from low mass galaxies (Forbes et al. 2018) and four clusters (Virgo, Coma, A1689, A2744) of galaxies (Harris et al. 2017). 
{ Like any other large sample study of GC counts, the sample used here includes a variety of observational approaches and analysis techniques. Typically, multi-filter images are taken of the host galaxy, GC candidates are selected and contaminants (i.e. background and foreground objects) removed statistically. For more distant galaxies, the imaging may only detect the brighter GCs. The total number of GCs is then estimated by counting those brighter than the (assumed symmetric) peak in the GC log-normal luminosity function and multiplying this number by two. 
This assumption is a good one for large, GC rich systems. For low mass galaxies that tend to be poorer in GCs (and for which the assumption of a symmetric GC luminosity function may break down), we mitigate this effect in two ways. First, we focus on deep imaging studies from HST which sample $\sim$90\% of the GC luminosity function at the distance of Virgo and Fornax, so that the correction to a total GC count is small. Secondly, for Local Group dwarf galaxies individual GCs are counted directly from imaging and should be a fairly complete representation of their GC systems (see Forbes et al. 2018). For the 
Eridanus, Fornax and Antila clusters we follow Spitler \& Forbes (2009) and increase the GC counts by 80\% to account for intracluster GCs and GCs around other cluster galaxies. }
The virial masses taken from Spitler \& Forbes (2009) are statistical measures from weak lensing, whereas for both, the Forbes et al. low mass galaxies and the Harris et al. clusters of galaxies, the virial masses are derived for the individual systems, based on their observed kinematics, X-rays etc. Our sample includes galaxies of all types, although it is dominated by early-type galaxies. 

We also include the recent observations of the Coma cluster ultra diffuse galaxy DF44 by van Dokkum et al. (2019). This is the only ultra diffuse galaxy with a published radial kinematic profile that extends beyond the half-light radius. van Dokkum et al. fit cusp and core dark matter models to the radial kinematics, finding that the core model provides a better fit. Their derived virial mass for the core model is log M$_{vir}$ = 11.2$\pm$0.6. The galaxy contains a substantial GC system with N$_{GC}$ = 74$\pm$18. 

\begin{figure*}
	\includegraphics[width=1.95\columnwidth]{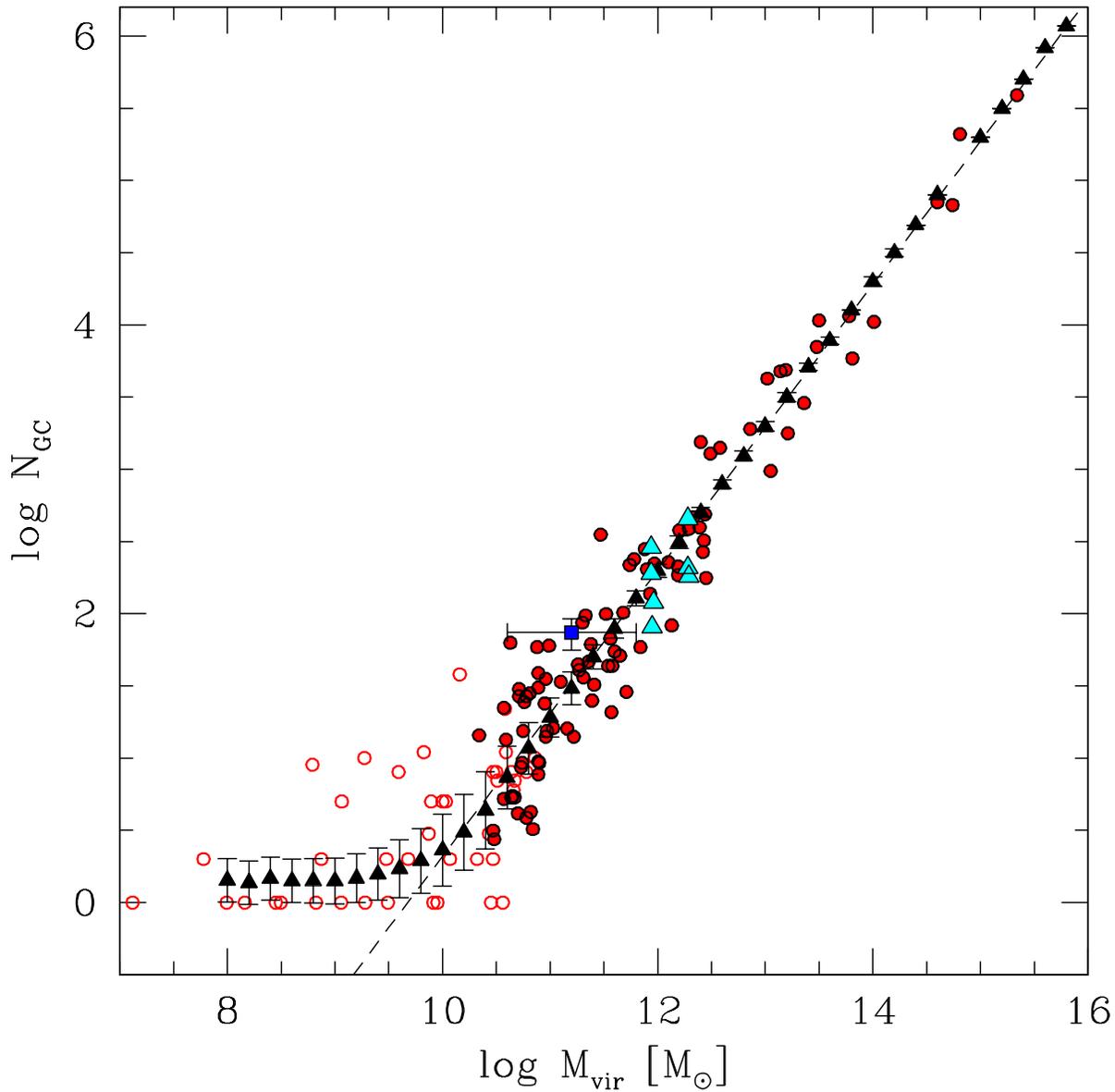}
    \caption{Observed correlation between the number of globular clusters $N_{GC}$ versus the virial mass of their host dark matter halos $M_{vir}$. The data extends over 7 magnitudes in mass from dwarf galaxies (open red points) to disk galaxies (cyan triangles) and ellipticals, all the way to galaxy clusters (both shown by filled red points). Only halos which contain at least one globular cluster are shown. The blue square with error bars shows the first ultra diffuse galaxy DF44 with a reliable determination of its halo virial mass. The dashed line depicts the empirical correlation (equation 1). The black triangles show a Monte Carlo merger simulation that starts with a seed halo population in the dwarf galaxy regime ($10^8 M_{\odot} \leq M_{vir} \leq 10^9 M_{\odot}$), filled randomly with 0,1 or 2 globular clusters, such that on average one globular cluster per $5 \times 10^9 M_{\odot}$ of dark matter is generated. In this simulation a new halo is formed in the seed regime after every merger event in order to keep the total number of halos constant. Error bars denote the variance around the mean of the merger population which strongly decreases towards higher masses due to the central limit theorem, in contrast with the observations.}
\label{fig:figure1}
\end{figure*}

The red points and cyan triangles in Fig.1 show the observations. The data extends from dwarf galaxies (red open points) with log(M$_{vir}$/M$_{\odot}$) $\geq$ 8 to clusters of galaxies with log (M$_{vir}$/M$_{\odot}$) $\sim$ 15, i.e. 7 orders of magnitude. A least-squares fit, neglecting the dwarf regime (log (M$_{vir}$/M$_{\odot}$) $\leq$ 10) where a correlation can only be valid statistically (see below) gives a perfectly linear relationship with:

\begin{equation}
\langle \log N_{GC} \rangle = -9.58 (\pm 1.58) +0.99 (\pm 0.13) \times \log M_{vir}/M_{\odot}
\end{equation}

\noindent This correlation implies a characteristic dark matter mass in virialized halos per globular cluster at z=0 of

\begin{equation}
M_{DM,GC} = \langle M_{vir}/N_{GC} \rangle = 5 \times 10^9 M_{\odot}.
\end{equation}

The correlation is independent of galaxy type. The cyan triangles in Fig.1 show that the 7 spiral galaxies in our sample lie perfectly within the distribution of early-type galaxies. More quantitatively, the spirals have mean virial masses of log $\langle$M$_{vir}$/M$_{\odot} \rangle $ = 12.09 $\pm$ 0.17 and GC numbers of $\langle$log N$_{GC} \rangle $ = 2.25 $\pm$ 0.23. For this virial mass, equation 1 predicts log N$_{GC} = 2.28$, in excellent agreement with the observations.  The blue square with error bars shows the first ultra diffuse galaxy (UDG) DF44 with a reliable determination of its halo virial mass (see van Dokkum et al. 2019 and references therein). Like all UDGs, this galaxy is extreme with a very extended stellar component, total mass M$_* \approx 3 \times 10^8$ M$_{\odot}$ within a half-light radius of 4.6 kpc and a kinematically estimated M$_{vir} \approx 1.6 \times 10^{11}$ M$_{\odot}$. Despite its strong deficit in stellar mass and its unusual extent, the observed GC number $N_{GC} = 74^{+18}_{-18}$ of DF44 is in excellent agreement with the linear correlation of equation 1. Note that equations 1 and 2 are valid for cosmic redshift z=0. Smooth accretion of dark matter implies that M$_{DM,GC}$ will decrease with increasing z, providing valuable insight into the growth and merger history of dark matter halos (see below). 

The linear correlation appears to break down in the dwarf galaxy regime for M$_{vir} \leq$ M$_{DM,GC}$ where galaxies typically contain a small number of GCs, if at all. This is however precisely what would be expected, if the equations (1) and (2) remain valid even in this mass regime. Figure 1 only shows galaxies that contain at least one globular cluster. For halos with masses below $\sim$ M$_{DM,GC}$, equation (2) can only be true statistically as in this regime, the fraction of halos that contain GCs  must be less than unity. For example, on average, only 1 out of 5 halos with M$_{vir} = 10^9$ M$_{\odot}$ should contain a globular cluster. In order to test this conjecture observationally one would need a statistically complete sample of dwarf galaxies with known N$_{GC}$ and M$_{vir}$, including especially galaxies without any globular cluster. This is not available yet. We will however show in the next section that the merging scenario provides insight into this question and indicates that equations (1) and (2) remain valid, statistically, down to virial masses of order $10^8$ M$_{\odot}$ (black triangles in Figure 1).

\section{The power of hierarchical merging}

A linear correlation between physical quantities usually points towards a fundamental and simple origin. Hirschmann et al. (2010) suggested hierarchical merging in order to explain the anti-hierarchical supermassive black hole (SMBH) growth and the linear correlation between central SMBH mass and galaxy bulge mass (e.g. H\"aring \& Rix 2004). A similar argument can also be applied to the excellent correlation between SMBH mass and the number of GCs (Burkert \& Tremaine 2010). Peng (2007) and Jahnke \& Maccio (2011) discussed near-linear scaling relations, in particular the SMBH--bulge mass relation, within the framework of the central limit theorem. We note, however, that SMBHs are not good tracers of the host galaxy virial mass (Kormendy \& Bender 2011). 
An example often quoted to illustrate this fact is the galaxy M33. It has a total GC system of 50 (Harris 2013) which would suggest a log virial mass of 11.39 according to our eq. 1, and yet it has no detectable SMBH to a limit of $<$1.5 $\times$ 10$^3$ M$_{\odot}$ (Gebhardt et al. 2001). 


Kruijssen (2015) and later on Boylan-Kolchin (2017) pointed out that an initially constant GC-to-halo mass ratio in a high-z galaxy seed population would be preserved during the hierarchical merging phase till z=0. El Badry et al. (2018) extended this analyses to an initial population with no correlation. Like Jahnke \& Maccio (2011), they argued that subsequent merging due to the central limit theorem would quickly establish a linear correlation and that this effect is so strong that the currently observed linear correlation for virial masses above $10^{11}$ M$_{\odot}$ does not contain any information about the physics of globular cluster formation and evolution. 

The situation is however more subtle. As a simple demonstration, let us start with a random merging model, following El Badry et al. (2018). Consider a seed population of n galaxies with a distribution of virial masses (M$_{vir,i}$: i=1...n) and globular cluster numbers (N$_{GC,i}$: i=1...n). Let us denote the average virial mass of this population as $\langle$M$_{vir} \rangle = \sum_{i=1}^n$M$_{vir,i}$/n and its average number of globular clusters as $\langle$N$_{GC} \rangle = \sum_{i=1}^n$N$_{GC,i}$/n. Following equation (2) the dark matter mass per GC is then M$_{DM,GC} = \langle$M$_{vir} \rangle / \langle$N$_{GC} \rangle$. Consider now one object that undergoes k mergers with other seed galaxies. For large values of k, its mass will approach a value of M$_{vir} \approx$ k $\times \langle$ M$_{vir} \rangle$, independent of its initial mass. The same is true for its globular cluster number which will approach a value of N$_{GC} \approx$ k $\times \langle$N$_{GC} \rangle$. In summary, a population of galaxies, growing by merging from a seed population will always move onto a universal linear correlation 

\begin{equation}
N_{GC} = \frac{\langle N_{GC} \rangle}{\langle M_{vir} \rangle} \times M_{vir} = \frac{M_{vir}}{M_{DM,GC}}
\end{equation}

\noindent with the location of an object on this correlation being determined by the number k of mergers. In reality the situation is more complex due to the fact that we also allow mergers between objects that had already experienced mergers. However these objects lie already close to the linear relationship. The merger remnant therefore just shifts along this relation to larger masses. Note that equation 3 contains valuable information on the initial seed population, i.e. its M$_{DM,GC}$ and by this on the physics of GC formation.

Fig. 2 shows a Monte Carlo simulation, similar to El Badry et al. (2018). We start with an initial seed population with virial mass, equally distributed in the logarithmic mass range $10.4 \leq$ log(M$_{vir}$/M$_{\odot}) \leq 11.1$, filled randomly with globular cluster systems with a total number of globulars being equally distributed in the logarithmic mass range 0.1 $\leq$ log N$_{GC} \leq 1.7$. We chose this dispersion in order to match the observed scatter of observations in the low-mass regime and to be consistent with equation 2. The box in the lower left corner in the upper panel of Fig. 2 shows the region initially populated with the seed population. We now select randomly two halos and merge them. The total virial mass and the total number of GCs after the merger is the sum of the virial masses and  GC numbers of each component and the total number of halos is reduced  by one. Repeating this procedure generates a distribution of halos with globular cluster systems as shown by the black points in the upper panel of Fig. 2. For comparison, the red points show the observations. In the lower left corner one can still see the seed population with no correlation while at the same time merging has produced a sequence of more massive halos that follow nicely a linear correlation. The observed linear trend is perfectly reproduced. We confirmed with additional test simulations that this result is independent of the initial distribution of seed virial masses and globular cluster numbers. It also does not depend on the average initial virial mass or average globular cluster number as long as the requirement is fulfilled that there is on average 1 GC per M$_{DM,GC} = 5 \times 10^9$ M$_{\odot}$ in the seed population.

\begin{figure}
	\includegraphics[width=1.\columnwidth]{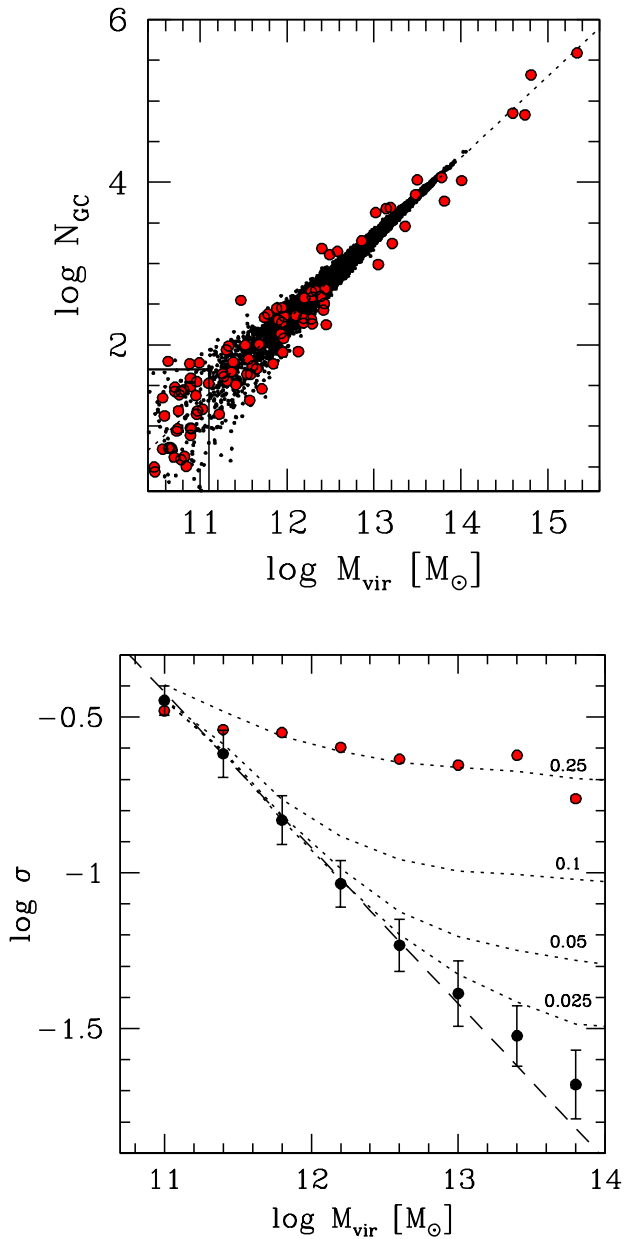}
    \caption{The {\it upper} panel show N$_{GC}$ versus M$_{vir}$ for dark matter halos that grew from a seed population (black points), randomly distributed within the lower left box. Red points represent the observations. The {\it lower} panel shows the logarithm of the scatter $\sigma$ of N$_{GC}$ around the mean correlation. Red points denote again the observations, black points the merger simulation. The scatter of the simulated GC systems around the mean decreases strongly as $\sigma \sim M_{vir}^{-1/2}$ (dashed line), in contrast with the observations which show a roughly constant $\sigma$. Dotted labeled lines show the dependence of $\sigma$ on errors $\delta$log(M$_{vir}$) in determining M$_{vir}$. Errors increase from 0.025 dex (lowest line) to 0.25 dex (uppermost line). The observations can be reproduced for $\delta$log(M$_{vir}$) = 0.25 dex, corresponding to an error of a factor of 1.8 in determining virial masses.}
    \label{fig:figure2}
\end{figure}

A second key physical property is the scatter of data points around the linear correlation. One can already clearly see in the upper panel of Fig. 2 that the variance in N$_{GC}$ in the simulation declines fast with increasing M$_{vir}$ while it remains roughly constant in the observed sample. This is quantified in the lower panel of Fig. 2 where we show the logarithm of the scatter $\sigma = \left( \delta N/N \right)_{GC}$ of the observations around the best fitting linear correlation (equation 1). For that, we first subtract the linear correlation $\langle$N$_{GC} \rangle$(M$_{vir,i}$) from every data point i with GC number N$_{GC,i}$ and virial mass M$_{vir,i}$ and then determine the scatter within a given logarithmic virial mass bin as the region containing 68\% of the data points. $\sigma$ is almost constant with a small decline for galaxy clusters. In contrast, the random merger model leads to a variance that declines as $\sigma \sim$M$_{vir}^{-1/2} $ (dashed line). This is a fundamental property of any merger scenario. The central limit theorem predicts that the scatter should change as $\delta$N$_{GC} \sim \sqrt{N_{GC}}$, leading to $\sigma_{GC} \sim \delta$N$_{GC}$/N$_{GC} \sim$N$_{GC}^{-1/2} \sim$M$_{vir}^{-1/2}$. As a result, this feature is always expected, independent of the adopted seed population or of the maximum virial halo mass where we stop the simulation. The strong decline of $\sigma$ therefore indicates that hierarchical merging should produce a very narrow correlation that allows a precise determination of halo masses, especially for massive halos with an accuracy of 

\begin{equation}
\delta \log M_{vir} \approx 0.04~ \rm{dex} \times \  (M_{vir}/10^{13}M_{\odot})^{-1/2}.
\end{equation}

At the moment, the observational data is still very limited. In order to explore the effect of the low-number statistics we took our merger simulation and randomly chose in each virial mass bin the same number of galaxies as observed. For this subsample we then determined again the variance. We repeated this calculation 1000 times and from that determined the scatter of $\sigma$ which is shown by the error bars in Fig. 2. They reflect the expected uncertainties in determining $\sigma$ for the case of a restricted sample of observed galaxies. The discrepancy between the observations and the simulation is still much larger than the error bars, demonstrating that one cannot explain this difference by low-number statistics. 

In the previous simulations we focused on the most simple case of random merging, with the number of halos decreasing by one with each merger. This affects especially the low-mass halos that get continuously depleted while more massive halos form. As a result, in the later phases of evolution the peak of the halo mass function shifts from the seed population to larger masses, approaching a Gaussian due to the central limit theorem. This is not in agreement with the dark halo mass distribution, inferred from cold-dark-matter simulations that follows a power-law mass function dN/dM $\sim$ M$^{-2}$ (Springel et al. 2008). In order to investigate the dependence on the halo mass function we performed Monte Carlo simulations where, after each merger event, we generated a new low-mass seed halo, following the same procedure as applied to the original seed population. Now the mass function of halos peaks at all times at the lowest halo mass. However, both, the emergence of a linear dependence and the strong decline in the variance of N$_{GC}$ proportional to M$_{vir}^{-1/2}$ is not affected, as long as the new population on average follows equation 2.

\section{Smooth cosmological accretion versus hierarchical merging}

The previous section neglected one key element of cosmological halo growth which is smooth accretion. {Smooth accretion, sometimes also called diffuse accretion, has to be clearly defined. Analysing halo merger trees, Fakhouri \& Ma (2009, see also Neistein \& Dekel 2008) find that the mass of a descendant halo does not completely match the sum of the masses of its resolved progenitor halos. This is a result of the infall of structures below their minimum resolved halo mass, accretion of dark matter that is not locked up in halos and mass loss due to tidal stripping.} In our context we define smooth accretion as the infall of dark matter particles that are not bound to substructure and of dark matter halos below a critical halo mass M$_{min}$ that could not seed any globular cluster, even statistically. A conservative estimate, obviously, is that halos with virial masses less than a GC mass could never form a globular cluster. M$_{min}$ could however be much larger than that. For example, Boylan-Kolchin (2017, see also Kravtsov \& Gnedin 2005) assumes that the formation of the blue (metal-poor) cluster population ended at z=6 with dark halos above a minimum critical mass M$_{vir} \geq$ M$_{min}$ hosting  N$_{GC}$ = M$_{vir}$/M$_{min}$ blue GCs. {By construction,} all substructures below this critical mass did not contain a GC. He then shows that one can reproduce the present-day correlation between blue GC-system mass and M$_{vir}$ if M$_{min} = 1.07 \times 10^9$ M$_{\odot}$. He also shows that for this value of M$_{min}$ smooth accretion dominates strongly the growth of dark halos. More precisely, the fraction $\eta_{smooth}$ of the present-day dark halo virial mass, resulting from smooth accretion is  $\eta_{smooth}$=89\%. As N$_{GC}$ = (1-$\eta_{smooth}$) M$_{vir}$/M$_{min}$ the linear correlation set up at high redshift will only be preserved if $\eta_{smooth}$ is independent of halo mass. Boylan-Kolchin (2017) indeed finds a constant, mass-independent value for massive galaxies. The situation changes however for smaller masses of log M$_{vir}$/M$_{\odot} \leq 10$, as discussed in the next section. {Note, that M$_{min}$, and correspondingly the value of $\eta_{smooth}$, is a free input parameter of the model. On the other hand, the fact that $\eta_{smooth}$ is independent of z=0 halo mass is a characteristic property of the scale-independence of cosmological structure formation and a key result of Boylan-Kolchin's analyses. Figure 3 (see also the discussion in Boylan-Kolchin 2017 and the discussion below) shows that the situation changes for z=0 halos with masses M $\leq 20 \times$M$_{min}$. In this regime the choice of M$_{min}$ begins to affect $\eta_{smooth}$ and, as a result, leads to deviations from the linear correlation between GC number and halo mass.}

\section{Extension into the dwarf galaxy regime}

In Fig. 1 the data points begin to deviate from the linear relationship for M$_{vir} \leq 10^{10}$ M$_{\odot}$ which is what one would expect if the rule of 1 GC per $5 \times 10^9$ M$_{\odot}$ extends all the way into the dwarf galaxy regime. As discussed previously, in this regime the correlation can only be valid in a statistical sense. To test this conjecture we ran another hierarchical merger simulation. Now we started with a seed population equally distributed in 8$\leq$log(M$_{vir}$/M$_{\odot}$)$\leq$9 which we populated randomly with 0, 1 or 2 GCs such that the population satisfies equation 2. The black triangles with error bars in Fig. 1 show the evolution of GC number with virial mass if we consider only halos that contain at least 1 GC. In excellent agreement with the observations we find a change in the linear correlation at $10^{10}$M$_{\odot}$, with a flat tail extending deeply into the dwarf galaxy regime.

\begin{figure}
	\includegraphics[width=0.95\columnwidth]{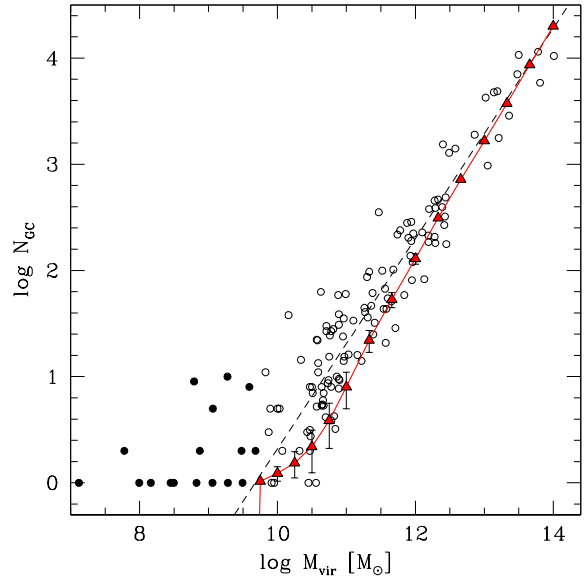}
    \caption{The observations (circles) are compared with the predictions of the Boylan-Kolchin (2017) model (red triangles with error bars). The model predicts no globular clusters in halos with $\log$ (M$_{vir}$/M$_{\odot}) \leq 9.5$, some of which are observed  (filled circles).}
    \label{fig:figure3}
\end{figure}

Figure 3 compares the Boylan-Kolchin (2017) model (red triangles, data kindly made available by Mike Boylan-Kolchin) with the observations (open and filled circles). Only halos with at least one GC have been considered. The model focuses on the blue GC subpopulation. In order to account also for the red (metal-rich) subpopulation we divide the GC numbers by the blue fraction f$_{blue}$, with (Harris et al. 2015)

\begin{equation}
f_{blue} = \left( \frac{M_{vir}}{10^{10}M_{\odot}} \right)^{-0.07}
\end{equation}

\noindent and the additional constraint that $f_{blue}$ is not allowed to drop below a value of 0.5 (high-mass regime with M$_{vir} \geq 10^{14}$ M$_{\odot}$) or above a value of 1 (low-mass regime with M$_{vir} \leq 10^{10}$ M$_{\odot}$). The simulations fit the data quite well, especially in the high-mass regime for virial masses $M_{vir} \geq 10^{12}M_{\odot}$. For smaller masses, the theoretical constraint of no GCs below M$_{min}=10^{10}$M$_{\odot}$ becomes visible and the correlation declines faster than linear. The correlation finally levels off close to M$_{min}$, in agreement with the observations. However, in contrast to the observations, the model predicts no globular clusters in halos below a virial mass of $5 \times 10^9$ M$_{\odot}$, that is the filled circles in Fig. 3 cannot be accounted for. Like the random merging model the scatter in N$_{GC}$, reflecting variations in the amount of smooth accretion, declines rapidly with increasing virial mass proportional to $M_{vir}^{-1/2}$ and becomes negligible for $\log$(M$_{vir}/M_{\odot}) \geq 12$. This confirms that cosmological merging leads to a very tight relation and that N$_{GC}$ is a precise tracer of dark halo mass, especially for massive halos. As argued below, the large scatter in the observations might then be dominated by errors in determining M$_{vir}$.

\subsection{Continuous GC formation and disruption}

What is the origin of the discrepancy in the scatter between observations and theory, shown in the lower panel of Fig. 2? Galaxies continuously {lose} globular clusters by disruptive processes (e.g. Fall \& Zhang 2001; Bose et al. 2018). If this is a random process, depending e.g. on the orbital parameters of GCs it should increase the dispersion of $N_{GC}$. In order to investigate this process we introduce a disruption parameter $d$ with $\log d$ equally distributed between 0 and $\log d_{max}$. Each time the virial mass of a galaxy in our merger model has grown by a factor of 2 we divide its number of GCs by $d$. Our merger simulations show that this process can indeed increase the scatter in $N_{GC}$ also for large values of $M_{vir}$. Good agreement with the observed scatter is however achieved only for very high values of $d_{max} \approx 6$ that would strongly affect the GC system.  In addition, the negative side effect now is that the correlation between $N_{GC}$ and $M_{vir}$ becomes shallower than observed. Of course one can enforce the linear correlation while increasing the scatter by generating on average as many GCs as are destroyed. Our merger simulation shows that for this, and adopting $d_{max} \approx 2.5$, one can indeed reproduce the observed spread, while at the same time maintaining a linear correlation. The agreement with observations can be improved even further if we adopt the requirement that no GCs are destroyed,  or form, for virial masses $\log(M_{vir}/M_{\odot}) > 13.6$. Although a viable model, the scenario of continuous disruption, balanced by continuous formation can be excluded. Globular clusters are on average very old systems (Strader et al. 2005). While GCs might continuously be disrupted by tidal stripping (Baumgardt \& Makino 2003) or by spiraling into galaxy centers, their old ages make it unlikely that as many GCs are continuously formed as are disrupted.

A likely ingredient of the observed scatter are uncertainties in virial mass determinations which all rely on tracers that sample certain halo regions with simplified assumptions about halo geometry and kinematics. The dashed lines in the lower panel of Fig. 2 show how the scatter $\sigma$ changes with virial mass for our hierarchical merger model if one adds a random, Gaussian error in determining M$_{vir}$ with a dispersion $\delta \log$ M$_{vir}$. For given logarithmic error, the decline in $\sigma$ flattens below a critical value which increasing with increasing $\delta \log$ M$_{vir}$. We find that one can reproduce the observations if $\delta \log$ M$_{vir} \approx 0.25$ dex, corresponding to uncertainties in virial mass determinations of a factor of 1.8. 

\begin{figure}
	\includegraphics[width=1.20\columnwidth,angle=-90]{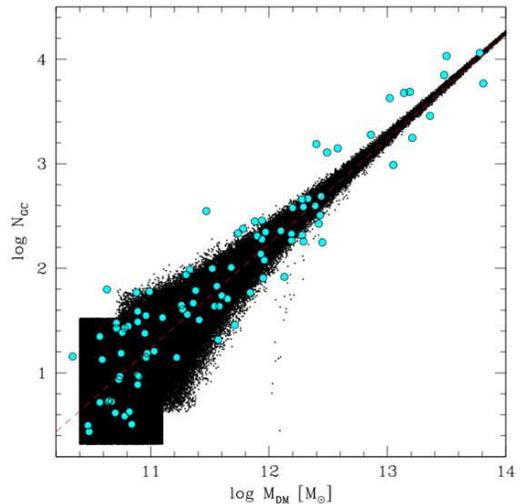}
    \caption{Observations (cyan circles) are compared with the hierarchical merger scenario (black points), taking into account a one-off GC system disruption event as soon as the galaxy has exceeded a critical mass of $10^{12}$ M$_{\odot}$. Besides a small tail of under-abundant systems close to that mass, there is no visible affect on the N$_{GC}$-M$_{vir}$ correlation.}
    \label{fig:figure4}
\end{figure}

There is no question that GCs are destroyed by tidal effects. A prominent example in the Milky Way is Pal 5 (Odenkirchen et al. 2002; K\"upper et al. 2015). If the disruption rate is independent of galactic environment it would not affect the linear correlation between N$_{GC}$ and M$_{vir}$, just M$_{DM,GC}$(z) which would then increase with decreasing redshift, not only because of an increase in the fraction of smooth dark matter accretion but also because of the loss of GCs as function of time. The situation becomes more interesting, if the GC disruption probability depends on the galaxy and dark matter mass. Mieske et al. (2014), for example, find that GCs should preferentially disrupt in galaxies with absolute visual magnitudes M$_V \approx$ --20 mag due to the fact that these galaxies have the largest 3D mass density within their half-light radius. The question then arises whether this mass-dependent disruption process would produce a visible signature in the correlation between N$_{GC}$ and M$_{vir}$, maybe even destroying the linear relationship. In order to test this question we studied an extreme case. Using our Monte Carlo approach we completely erased the GC system of a galaxy as soon as its halo mass exceeded $10^{12}$ M$_{\odot}$ by setting N$_{GC} = 0$. This was done exactly once during the evolution of each galaxy. Subsequent mergers could re-establish a GC system. Note that the precise mass where this happens varies as it depends on the mass of the last merger. Figure 4 shows the resulting correlation. Remarkably, this certainly drastic impact on the GC system has almost no visible affect on the evolution of the correlation. At the critical mass of GC disruption one now can see a tail of systems with a GC deficit. But hierarchical merging quickly erases this signature again. Interestingly, the cyan points (in Figure 4) show a very similar signature of a small group of under-abundant galaxies in this virial mass regime.

\section{Discussion and Conclusions}

The observations, collected in this paper, reveal a perfectly linear correlation between GC number and dark halo virial mass: we find that halos contain 5 $\times 10^9$ M$_{\odot}$ of dark matter per GC. We show that the correlation continues statistically into the dwarf galaxy regime, extending it to a mass range of 7 orders of magnitude. This empirical relation provides a powerful tool to determine halo virial masses, independent of any kinematic tracers or other methods, like weak lensing. Compared to previous discussions that focused on GC system mass, (e.g. Harris et al. 2017; Forbes et al. 2018) GC number has the advantage of not suffering from uncertainties in determining the luminosity L and assuming a mass-to-light ratio M/L for each globular cluster. One might argue that GC masses are dominated by massive GCs which are easily detectable while GC numbers are sensitive to low-mass GCs that are hard to find. However, in this analyses N$_{GC}$ was in general determined by counting the number of GCs above the peak in the luminosity function and multiplying this number by a factor of 2. It would therefore be even more accurate to argue that each GC above the peak in the GCLF contributes $10^{10}$M$_{\odot}$ of dark matter to the halo virial mass. With this procedure it should, in general, be easy to determine N$_{GC}$, as long as the peak of the luminosity function is well  resolved. We note that this method also makes N$_{GC}$ insensitive to secular disruption processes that affect preferentially the low-mass end of the GC luminosity function. 

The N$_{GC}$--M$_{vir}$ correlation contains important constraints for models of GC formation and halo growth (e.g. Adamo et al. 2015). We confirm previous claims that random merging is an enormously robust mechanism to produce the observed linear correlation. Taking into account cosmological smooth dark matter accretion does not change this result as long as the fraction of smooth dark matter growth is independent of final halo mass. Disruption of GCs would also not affect the correlation if the rate of disruption is independent of galaxy virial mass. It would however increase the value of the dark matter mass per globular cluster,  M$_{DM,GC}$. The central limit theorem is also remarkably efficient in erasing any signature of galaxy mass-dependent GC disruption (Mieske et al. 2014; Kruijssen 2015; Pfeffer et al. 2018). We demonstrate this with an extreme scenario, where all GCs are destroyed in a singular event, once a galaxy passes a critical virial mass. The result of this experiment is a weakly populated tail of galaxies with under-abundant GC systems, close to the adopted critical mass. The linear correlation is however quickly restored afterwards.

The Boylan-Kolchin (2017) model, where at cosmic redshift z=6 GCs populated halos with virial masses above M$_{min} = 10^9$M$_{\odot}$ is in good agreement with the observations of z=0 galaxies for virial masses above $10^{10}$M$_{\odot}$. No GCs should however exist in halos below this mass limit. This is in conflict with the detection of GCs in some dwarf galaxy halos with masses extending down to $10^8$M$_{\odot}$ (e.g. Forbes et al. 2018). Using our hierarchical merger model we had argued that these observations are still consistent with the linear correlation (equation 2), however only in a statistical sense. This would require a revision of the Boylan-Kolchin model which will be explored in a subsequent paper. Another possibility could be that these dwarfs with GCs actually formed in halos that were much more massive than usually assumed and that lost a large fraction of their dark matter mass e.g. by tidal stripping once they entered the potential of their host galaxy (e.g. Khalaj \& Baumgardt 2016). In this case, the Fornax dwarf spheroidal with 5 GCs, for example, would have formed in a halo with $2.5 \times 10^{10}$M$_{\odot}$ which is a factor of 25 more massive than its currently estimated virial mass of $1.1 \times 10^9$M$_{\odot}$ (Forbes et al. 2018). Interestingly, such a large virial mass would be consistent with abundance matching estimates (Moster et al. 2013) for Fornax's stellar mass of $4 \times 10^7$M$_{\odot}$ (de Boer et al. 2012).

A new important quantity that has not been explored so far is the scatter of data points around the linear relationship. We find that random merging necessarily produces a dispersion in $\log N_{GC}$ that decreases proportional to $M_{vir}^{-1/2}$. The same is true for the Boylan-Kolchin cosmological model, including smooth accretion. This first of all implies a very tight correlation for more massive halos, making equation 1 or 2 a precise estimator of virial masses. Secondly, it is in contrast with observations that indicate a scatter that is largely independent of M$_{vir}$. This discrepancy can be used in order to determine current uncertainties in deriving virial masses with kinematic tracers or weak lensing. We find that an error of a factor of 1.8 can explain the scatter in the observations.

The hierarchical merger model of Boylan-Kolchin (2017) assumes GCs to form in small substructures at high cosmic redshifts of $z \geq 6$. Strictly speaking, this model applies only to the blue, metal-poor subpopulation. Forbes \& Remus (2018) indeed find that cosmological models are consistent with the blue GCs to have formed largely ex-situ of present-day galaxies while the red GCs formed largely in-situ in massive galaxies. The linear correlation of equation 1 was however a result of looking at the total population (for a breakdown into red and blue GCs see Harris et al. 2015 and equation 5). This requires 

\begin{equation}
M_{vir}/N_{GC} = M_{min} \times \frac{f_{blue}}{1-\eta_{smooth}} = 5 \times 10^9 M_{\odot}.
\end{equation}

\noindent The most simple explanation for a constant M$_{vir}$/N$_{GC}$ would be a constant f$_{blue}$, in addition to a constant $\eta_{smooth}$. The ratio of blue-to-red GCs is indeed constant to good approximation for high masses. The situation is however different for smaller masses (equation 5). The red subpopulation might therefore also have formed ex-situ and early in lower-mass galaxies which later on continued to merge into larger structures. This merger process lead to a gradual change in f$_{blue}$ while at the same time establishing the linear correlation between M$_{vir}$ and N$_{GC}$. 

In order to distinguish between the various scenarios of GC formation and to better understanding the emergence of the correlation M$_{vir} = 5 \times 10^9$ M$_{\odot} \times$ N$_{GC}$ observations of GC systems as function of cosmic time will be essential. They might answer several key questions. Is the linear relation preserved all the way to high redshifts $z \approx 6$? How does f$_{blue}$ as function of M$_{vir}$ evolve with time?  What is the redshift dependence of the zero point M$_{DM,GC}$ (equation 2) which provides important insight into the evolution of the fraction of smooth halo accretion $\eta_{smooth}$? Answers to these questions will also give valuable insight into cosmic structure formation and galaxy assembly in general.


\section*{Acknowledgements}

We thank M. Boylan-Kolchin, W. Harris, J. Kormendy, D. Kruijssen, B. Moster and S. White for inspiring discussions { and the referee for constructive and important suggestions}.  
Thanks to M. Boylan-Kolchin also for providing the data of his cosmological merger model. AB and DAF thank the department of astronomy at the University of California, Santa Cruz, for their hospitality during the preparation of this paper. 
DAF thanks the ARC for financial support via DP160101608.

\acknowledgments

\end{document}